\documentclass[intlimits,twoside,a4paper]{article}

\usepackage{amsmath,amssymb}
\usepackage{graphicx}

\usepackage[cp1251]{inputenc}

\usepackage[eqsecnum]{cmpj3}

\usepackage{color}
\newcommand{\tern} {Cu$_{60}$Ti$_{20}$Zr$_{20}$}
\newcommand{\PP}{PP}

\issue{2017}{20}{3}{33702}
\doinumber{10.5488/CMP.20.33702}

\title[Liquid metals]
{Liquid metals: early contributions and some recent developments}
\author{C. Regnaut, S. Amokrane}
\address{Groupe ``Physique des Liquides et Milieux Complexes'',  Facult\'{e} des Sciences et
Technologie, Universit\'{e} Paris-Est (Cr\'{e}teil), 61 av. du G\'{e}n\'{e}ral de Gaulle,
94010 Cr\'{e}teil Cedex, France }

\date{Received May 11, 2017, in final form June 29, 2017}
\begin{document}

\maketitle

\begin{abstract}
We illustrate  in this contribution the progress in the theoretical study of liquid metals made in the last decades, starting from  the example of liquid gallium  and  the early work in Jean-Pierre Badiali's group. This was based on the combination of the perturbation theory with  pseudo-potentials for the electrons and the liquid state theory for the ions. More recent developments combining \textit{ab initio}  and  classical molecular dynamics simulations  are finally illustrated on the example of glass forming alloys.

\keywords liquid metals, structure, perturbation theory, simulation
\pacs 71.22.+i, 61.20.Gy, 61.20.Ja, 71.15.Pd
\end{abstract}

\section{Introduction}

Among  various fields of research in which Jean Pierre Badiali (JPB) was interested in, we want to emphasize in this contribution some of his early  works on liquid metals  and discuss some recent developments in this field.  In the seventies, Jean Pierre  actually devoted  a large part of  his research to the application  of  statistical mechanics to the electrochemical interface. He wanted to reach a description of the metal/solution interface at the atomic level  in order to develop a  multiscale approach of  reactivity at the  interface. This effort   illustrated by an early  study of the current-voltage relationship \cite{JPB72} culminated in a series of papers on the contribution of the metal to the properties of the interface, ranging from the capacitance of the ideally polarized electrode \cite{JPB75, MLR82,SA89,SA92} to the short-range interactions at electrodes coated with conducting polymer films \cite{Vorotyntsev94}. In the mean time,  JPB was  convinced that an important task was to accurately describe the bulk solution and  the bulk metal by using the best available theories, an effort which should be beneficial to the modelling of the interface itself. In the 1970s,  the theory of liquids and metals  was rapidly growing. As concerns  the  liquid  state,  integral equations, perturbation theories and variational approaches  gave a new impulse to the understanding of  simple  liquids and their mixtures, thanks to the success of the hard sphere (HS) model as a reference system \cite{Hansen}.  Regarding the metal studies, the pseudopotential (\PP)  theory (model or \textit{ab initio}) was rapidly developing after the first studies  in the 1960s  \cite{Harrison}, allowing one to derive an effective pair potential between the screened ions or pseudoatoms in the metal. For these reasons,  JPB decided to use this approach to  study the bulk liquid metal. Since we have  pursued until nowadays  this  early work impulsed  by JPB, in different contexts, we  chose to focus our contribution on these metal studies, including some developments  beyond the pair potential approach, that were not present in JPB's work.  Actually, liquid and amorphous metals have been the subject of an intense theoretical and experimental activity for decades now, up to forming or being among  the major themes of a series of international conferences that started in the mid-sixties\footnote{ Liquid Metals conferences: LM1, Brookhaven, 1966; LAM16, Bonn-Bad Godesberg, 2016.} and early-nineties\footnote{Liquid Matter conferences: LMC1, Lyon, 1990; LMC10, Ljubljana, 2017.}. Understandably,  an overview of the related literature would be a very difficult  task and  we shall not attempt to do this here --- see for instance the summaries up to 2002 in \cite{Cowlam,Hensel}.

Although mercury or noble metals were more often used  in electrochemical cells, JPB focused first on gallium  because this metal was extensively studied in the group\footnote{Laboratoire de Physique des Liquides et Electrochimie, University Paris 6, France.} \cite{Badiali74}  and, among all simple metals (i.e., $s$-$p$ bonded ones), was  very  puzzling for theorists. Indeed, when compared with other trivalent metallic elements, Ga  appears  very strange: low melting point temperature $T_{\text M}$, exceptional polymorphism and metastability, anomalous liquid structure factor and  huge supercooled  liquid range  of  about $- 0.5 T_{\text M}$ \cite{Defrain,Bizid, Badiali80}. Thus, gallium was, at a first glance, a severe test  both for pseudopotential and liquid state theories. After a brief overview of the theoretical methods (section~\ref{sec2}), section~\ref{sec3} of this paper is thus devoted to pointing out some of  the results  early obtained with JBP,  to explain qualitatively the structure of liquid Ga  (l-Ga)  and justify the relative stability of its  numerous solid phases. Then, we  briefly survey the state of the art in gallium studies in the 2010s, especially  the considerable  progress made since the advent of the so-called \textit{ab initio} molecular dynamics (AIMD) methods, as contrasted with classical molecular dynamics (CMD) with parametrized force fields.  As a simple illustration, an example of our own AIMD  calculations of the  l-Ga structure will be given.
After gallium, JPB  focussed in the 1980s on the noble metals. With his co-workers, he used the best  available models, i.e., the resonant model \cite{Dagens,Regnaut83}  or  the generalized \PP\  theory (GPT) of Moriarty \cite{Moriarty,Moriarty22,Moriarty33}  to include the full $d$-band, the $s$-$d$ hybridization  and overlap contributions to derive the effective atomic interactions in noble metals and the corresponding  liquid structure and properties \cite{Regnaut85}.  In section~\ref{sec4}, we  briefly summarize some of these  results.  Beyond this, the orbital free AIMD method  for liquid noble metals  \cite{Bhuiyan}, will be mentioned.  Finally,  in connection with the work of JPB and coll.  for copper, we show in subsection~\ref{sec5} the recent AIMD results on Cu based alloys  by Amokrane et al. \cite{SA2015} in a study that illustrates the efficiency of some modern methods combining AIMD and CMD on the non-trivial  case of the glass forming ternary alloy \tern, a task hardly conceivable with the methods used 40 years ago.

\section{ Methodological aspects}\label{sec2}

\subsection{The pair potential approach}
Historically, the modern theory of metals emerged from quantum mechanics with the development of interatomic potentials. This was done first in the case of nearly free electron metals for which the electron-ions pseudo potential is sufficiently weak to be treated by perturbation expansions \cite{Harrison,Cohen_Heine,Ashcroft_Langreth}. During nearly thirty years, an intense activity took place to develop interatomic potentials  using either  first principles \PP s or simple local models and to extend the perturbation treatment  for $s$-$p$ metals (see for example \cite{Shaw,Shaw22,Ashcroft} and \cite{Hafner87}  for a review) to transition metals, either with full $d$-bands or those with partially field $d$-bands (see  \cite{Moriarty,Moriarty22,Moriarty33}).  The early work in JPB's group was done in this general framework.

For a given pair potential, the liquid state properties were  obtained from the methods of statistical mechanics \cite{Hansen}, initially mostly perturbation theory or integral equations for the pair structure,  such as the  optimized random-phase approximation \cite{Andersen}  (ORPA)  or the modified hypernetted chain (MHNC) \cite{Lado,Rosenfeld}, see for example \cite{Regnaut79,Kahl_Hafner87,Kahl_1990}.  Later on, the development of the computing power led to a  gradual shift from integral equations to simulation, mostly molecular dynamics, see for instance \cite{Hoshino,Canales,Wax}.
In parallel with the efforts to improve the interatomic potentials, possibly including many-body contributions, the  progress made in the calculation of the electronic structure  by  DFT and a dramatic increase in the computing power led in the early nineties  to a radically different approach, as detailed below.

\subsection{Towards the potential energy surface }

The mid-eighties witnessed the development of efficient algorithms for computing the electronic energy for a large number of ionic degrees of freedom, as well as a dramatic increase of the computing power of affordable computers.  The effort switched then from  the development of  semi-analytic first-principles interatomic potentials to the purely numerical treatment of the  full quantum mechanical manybody electron-ion system. Figure~1 in Moriarty's \cite{Moriarty2006} paper in 2006  provides a clear  graphical  display of the new situation.
The electronic structure is treated in the framework of the  Kohn-Sham \cite{Kohn-Sham} density functional theory (DFT), in combination either with  an extended lagrangian formulation as in the Car Parinello method \cite{Car} or by  directly computing the forces from the ground state potential energy surface $E(\textbf{R}_1,\ldots,\textbf{R}_N)$ for each instantaneous  configuration of the $N$ ions ---  see \cite{Martin} and references therein for the basis of  this Born-Oppenheimer molecular dynamics (BOMD) and \cite{Kresse93,Kresse932,Kresse-Furthmuller}  for actual implementations. The accuracy of these methods benefited from the advent of improved \PP s, among which the projector augmented wave (PAW) \PP\ \cite{Blochl,Kresse-Joubert} can be as accurate as  full-potential approaches \cite{Martin,Holzwarth}, such as  the LAPW one \cite{Singh}.  They presently form the basis of  AIMD simulations in condensed matter physics and materials science.  Starting in the early nineties, they have been implemented by different academic groups.  Among the various implementations,  the self-consistent plane wave  (PWscf) fortran code in the open source Quantum-Espresso package  \cite{QE} has been used in this work and in \cite{SA2015}.

 In the AIMD of extended systems, pseudo-potentials are still required to avoid   costly all-electrons calculations but they are used differently compared to in the interatomic potentials approach, since the DFT treatment of the electronic structure in the total external field created by the ions  is not perturbative.  According to the review of Kresse in 2002 \cite{Kresse2002} and references therein,  most of the metallic  elements can be fairly well described using  PAW pseudopotentials. The  large number of electrons one needs to consider with accurate PAW \PP s,  limits, however, the size of the  systems that can actually be simulated without large computer resources.
 The remaining limitations are mostly observed for the low melting point elements.  Other limitations also occur due to the approximation made in the DFT such as the local density approximation (LDA) and generalized gradient approximations (GGA).

 As a first step, one may also use the  orbital free variant of DFT in which only the density is required to evaluate the electronic energy functional.
Like  for  the simple metals, this  considerably improved the description of liquid noble metals (see, for instance, the work by Bhuiyan et al.  \cite{Bhuiyan} and references therein).   The full Kohn-Sham (i.e., with orbitals)  calculations are in principle more accurate.  However, even with modern computers, the AIMD  simulations  become  impractical if they are to be repeated at different points in the parameters space, as, for example, in the construction of  phase diagrams --- see e.g., \cite{Harvey} and references therein. One  needs then  to resort  to classical molecular dynamics, with parametrized force fields. (See, for example,  \cite{Alemany} for transition metals.  Embedded atom  models (EAM) potentials for fcc elements are given, for example, in \cite{Sheng} and $n$-body potentials for ternary alloys are reviewed in \cite{Lia}.) In the last section, we thus  illustrate the issue  of the appropriate combination of \textit{ab initio} and classical MD on the example of our own work on glass-forming transition metals alloys.

\section{ Gallium: from pseudopotentials  to \textit{ab initio} molecular dynamics}\label{sec3}

As just mentioned, gallium is known to have numerous structural and thermodynamical peculiarities. The salient ones in the liquid state are: a  low melting temperature $T_{\text M} = 303$~K; a density higher than in the stable solid phase; a very large supercooling range of about 150~K; under 257~K and normal pressure, the supercooled liquid  prefers to crystallize in the metastable $\beta$-Ga form  \cite{Defrain}; finally, its non-HS-like liquid structure factor with a shoulder in the main peak. Therefore, we first paid attention \cite{Badiali74}  in 1974 to the difference between a  HS-like structure  factor $S_{\text R}(q)$ as a reference for gallium and the true one $S(q)$.  From the \PP\ theory  and using the random phase  approximation, a very simple expression connects  at any temperature  this structural difference $\Delta S(q)$  with $\Delta V(q)$,  the corresponding difference  in $q$-space between the effective interionic interaction and the reference one.  For  a local  \PP, the relative change $\Delta S(q)/S(q)$ is linked to the change in the  \PP\  by \cite{Badiali74}:
\begin{equation}
\frac{\Delta S(q)}{S(q)} = \frac{S_{\text R}(q)\Delta V(q)}{kT} \approx-\frac{1}{kT} S_{\text R}(q)f(q)[w^2 (q)-w_{\text R}^2 (q)],
\label{delsq}		
\end{equation}
where $f(q)$ includes the standard  functions of the perturbation theory for nearly free electrons \cite{Harrison}. This function that depends only on the  Fermi wavenumber $k_{\text F}$ and the valence,  induces the so-called Friedel oscillations  of $V(r)$ at a large distance. In equation~(\ref{delsq}),  the Fourier transform of Ashcroft's empty  core model \cite{Ashcroft} (ECM) is $ w(q)= -\frac{4\piup Z}{ \Omega q^2} \cos(qr_{\text c})$, where $\Omega$ is the atomic volume in the liquid, $Z$ is the valence (or effective one) and $r_{\text c}$ is the core radius. This approximate expression of $\Delta S(q)/S(q) $ highlights  the three pertinent parameters $Z$, $k_{\text F}$ and $r_{\text c}$  governing the shape of $S(q)$ and its change from one metal to another. Moreover, since  l-Ga (non-HS-like) and aluminium (HS-like) have the same nominal $Z$ and  almost the  same  $k_{\text F}$ value, the only difference between these two metals lies in the core radius $r_{\text c}$. Unfortunately, we estimated with JPB  that $w(q)$ should be known with an accuracy better than 5\%,  while the differences  found when using  various \PP s and expressions of the screening function involved in $f(q)$ are much larger. Therefore, it was concluded that if the important structural  and thermodynamic features of l-Ga lie first in its specific \PP, both non-locality and  accurate  liquid state theory methods  are required. Using one of the best available non-local model potential developed by Shaw \cite{Shaw,Shaw22} to obtain $V(r)$ and the ORPA  to obtain the liquid structure factor,  it was  shown that Shaw's model can predict the right trends in the structure factors of Mg and Al (HS-like) and Zn, Ga, Sn (non-HS-like) \cite{Regnaut79}, but the  relative stability of the principal solid phases of Ga could not be  fully established without some ad hoc empirical adjustments \cite{Regnaut80}. On the same basis but with the ECM  and ORPA, Hafner and Kahl studied later on the structural trends of the metallic elements in the liquid state over the periodic table in a 3D space \cite{Hafner_Kahl} (using $Z$, $k_{\text F}$ and $r_{\text c}$ as  variables), following a parent study  for their solid structure \cite{Hafner83}. They concluded  that the complex structure of the light polyvalent liquid metals arise from the interplay of two characteristic distances: the effective diameter $\sigma_{\text{eff}}$ of the  repulsive core of $V(r)$ and the  wavelength of the Friedel oscillations in  the potentials, in full agreement with the studies of JPB and coll.

Unfortunately, other studies and many attempts with different \textit{ab initio} or models  \PP s indicate that gallium remains a troublesome metal. Indeed,  among the liquid polyvalents metals, some of them although non-HS-like (for instance Ge) are very well described either from the one parameter ECM (Hafner and Kahl \cite{Hafner_Kahl}) or Shaw's parameter-free model \cite{Regnaut88}.  By contrast,  Ga  is then poorly described,  the ECM being incapable of predicting the stability of the solid form GaI, while  Shaw's one does this qualitatively \cite{Dupont_These,Regnaut_These}.

A few years later, the AIMD methods raised new hopes for accurately describing  gallium and non-simple metals, although the LD and GG approximations  seem to have an influence on  determining precisely the structure of liquid gallium \cite{Kresse2002}, as well as the relative  stability of its crystalline structures at normal and high pressure, at the observed equilibrium volumes \cite{Bernasconi}. Indeed, previous studies with OPW pseudopotentials \cite{Hafner75} can better predict  the equilibrium volumes. However, describing all the  metastable crystalline $\beta$, $\gamma$ and $\delta$   phases  remains  a considerable challenge since a full agreement can be obtained only after very small empirical corrections \cite{Dupont_These,Regnaut80}.

The most recent AIMD simulation of  liquid gallium we are aware of is that of Chen et al. \cite{Chen2016}. The authors simulated a system of 1331 Ga atoms at $T=323$~K at the experimental density $\rho=0.0525$~\AA$^{-3}$, using a PAW \PP \ and the LDA. They analyzed the bond-orientational order with the aim of identifying the possible source of  a shoulder in the first peak of $S(q)$. They  concluded that  clusters, especially those with fourfold symmetry,  could explain this  special feature, without excluding the interpretation based on the Friedel oscillations.

\begin{figure}[!t]
\centering
\begin{minipage}{0.49\textwidth}
\begin{center}
\includegraphics[width=0.99\textwidth]{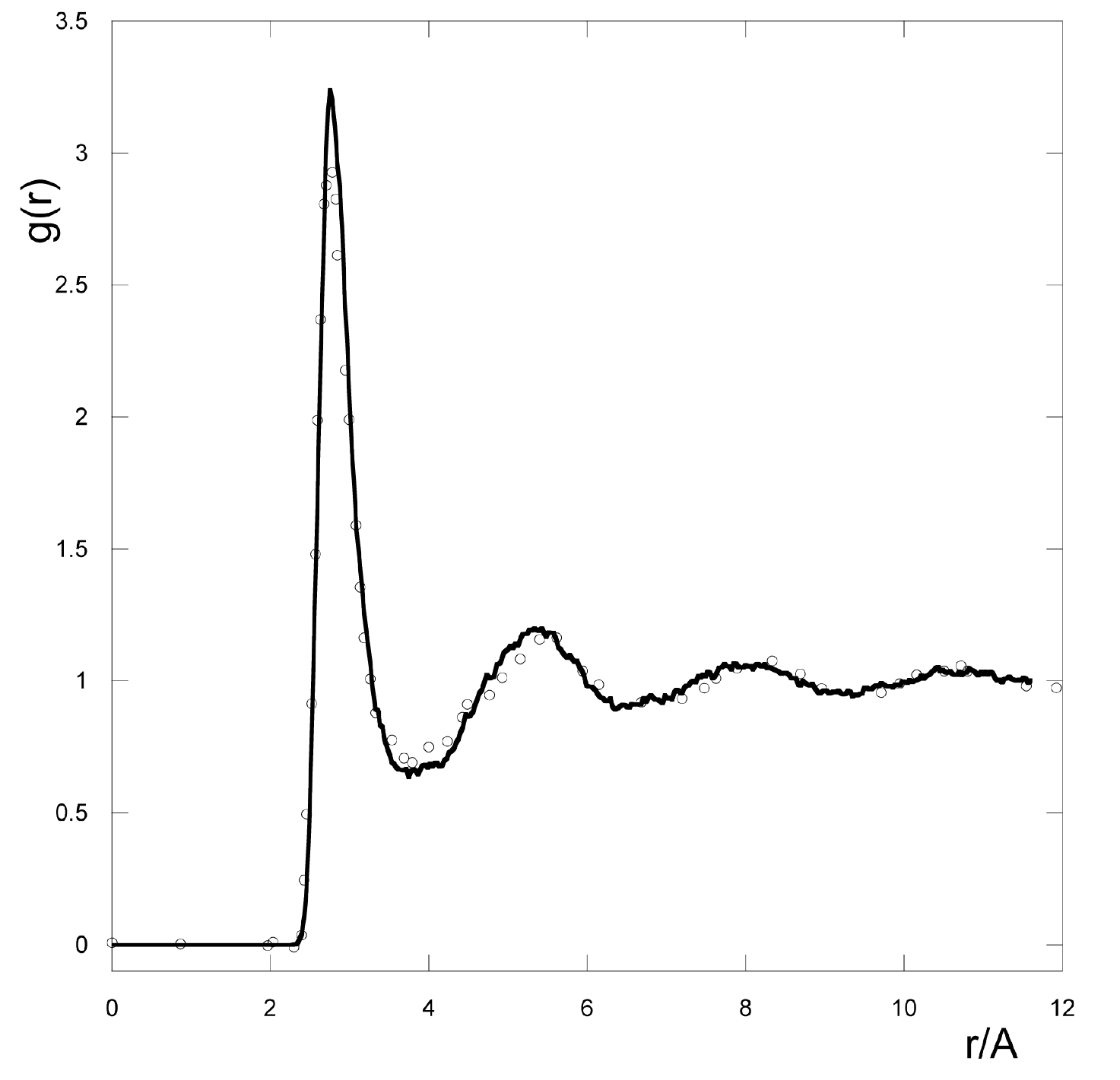}
\end{center}
\end{minipage}
\begin{minipage}{0.49\textwidth}
\begin{center}
\includegraphics[width=0.99\textwidth]{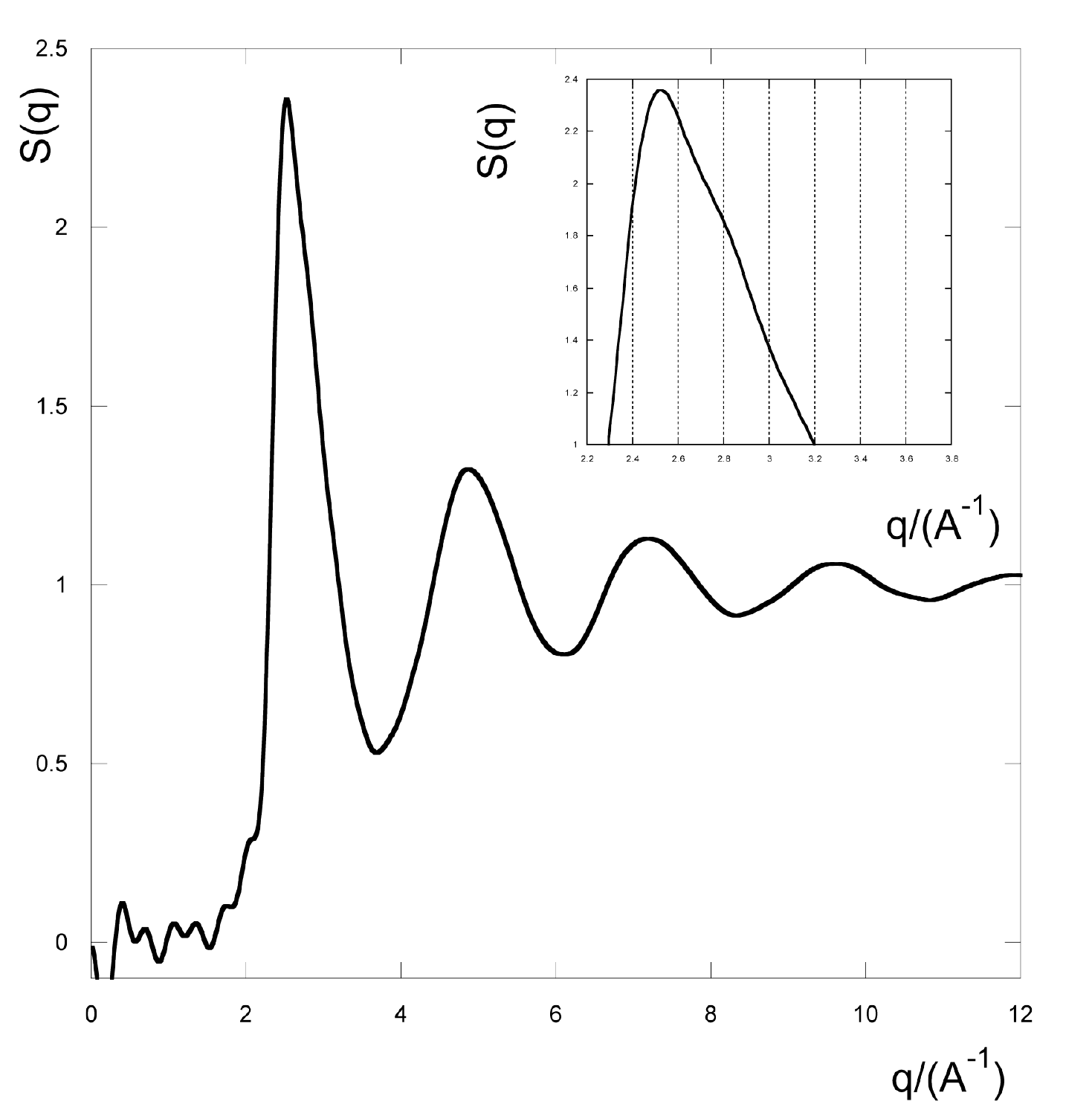}
\end{center}
\end{minipage}
\begin{minipage}{0.49\textwidth}
\vspace{-4mm}
\begin{center}
\caption{\label{gr_Ga} Experimental and theoretical  radial distribution function of liquid Ga at T$=323$~K. 
Symbols: xrd data \protect \cite{Bizid}, full curve: AIMD, this work.}
\end{center}
\end{minipage}
\begin{minipage}{0.49\textwidth}
\begin{center}
\caption{Structure factor of liquid Ga at T $=323$~K.
$S(q)$ is computed from the Fourier transform of the AIMD $g(r)$ in figure~\ref{gr_Ga}. Inset shows the first peak on an enlarged scale.}
   \label{SqGa323N256}
\end{center}
\end{minipage}
\end{figure}

\looseness=-1 Shorter  calculations may, however, be useful for a rapid estimation of the pair structure. To illustrate this,  we performed a simulation with $N=256$ and a norm conserving (NC) \PP\ with only  $Z = 3$~electrons  (one MD step lasts about 130~s on our 24~cores machine). The total number of electrons  is  thus 22.5~times smaller than in a simulation  with 1331~ions and $Z=13$ with the available PAW \PP s (recall that the computation time increases much faster than the number of electrons). Figure~\ref{gr_Ga} obtained after 1000~steps of equilibration and 1000~steps of accumulation illustrates this possibility to determine a reasonable $g(r)$ with moderate resources (x-ray diffraction (xrd)  data are shown for comparison).  Partial results after  650~MD steps with a PAW \PP\ showed no significant difference, but this conclusion needs to be confirmed --- one MD steps with $N=256$ is about 17~times longer than with the NC   \PP. The corresponding structure factor shown in figure~\ref{SqGa323N256} obtained by Fourier transform of $g(r)$ shows a clear asymmetry of the first peak, though the shoulder is less resolved than in figure~2 of \cite{Chen2016} or in the pair potential approach. Among the possible reasons to this, the small system size, the short runs and  simple \PP\ used here to speed up the calculation.

Note that in a previous AIMD study on l-Ga under pressure \cite{Yang},  the analysis of the local geometry and dynamical behavior refutes the hypothesis of Ga$_{2}$  dimers in the liquid state. It  reinforces the  interpretation of the shoulder in $S(q)$ for  l-Ga as  a result of medium-range order beyond the first shell caused by Friedel oscillations. The authors concluded that their results were consistent with experimental studies \cite{Ascarelli,Nield} in which no evidence was found  for any type of ``molecular'' species at a distance like that in $\alpha$-Ga, nor for any specific type of clustering \cite{Nield}. This conclusion was drawn from the analysis of the pair structure. A detailed analysis of the local structure,  however,  requires going beyond the distribution of the distances, as in \cite{Chen2016}.

A similar interpretation based on some medium-range order  related to the structures beyond the first shell in the radial distribution function (rdf) caused by a cooperation between the short repulsive core and the Friedel oscillations in the pair potential has been discussed in \cite{Ten-Ming}. The situation seems more contrasted in the case of germanium \cite{Takeuchi}.

\looseness=-1 In summary, AIMD simulations provide a quantitative description of the pair structure of l-Ga, including a shoulder in the principal peak of $S(q)$,  that the pair potential approach is able to reproduce  in an effective way.  Nonetheless, simulations with large systems are required to definitely assess the precise features in the local structure which are  responsible for these collective features, as in \cite{Chen2016}. Moreover,  AIMD studies in the deep undercooled states are particularly demanding since they require large enough systems to resolve transient structures and a sufficient  number of time steps, due to  the corresponding slow dynamics.

\section{Beyond simple metals and the interatomic potentials approach}\label{sec4}
In this section, we summarize some other results  concerning non-simple metals and alloys.  Again, we shall not attempt a review of the literature,  focusing mostly on the results obtained following JPB's  early work.  We next present  some  of our recent extensions of this work.
\subsection{Noble and transition metals}

As  underlined in the introduction, JPB's interest  in noble metals  was motivated by  their importance  in electrochemistry, in particular, as electrode materials. Transition  metals are actually much more difficult to study from first principles pseudopotentials than $s$-$p$ bonded ones due to the $d$-band contribution.  However, the noble metals Cu, Ag, Au,  have full $d$-band and the theoretical predictions are generally better than for metals having a partially filled  $d$-band.  In the 1980s,  the  best  available approaches  for noble metals were  the resonant model \cite{Dagens} and the generalized pseudopotential theory (GPT) of Moriarty \cite{Moriarty,Moriarty22,Moriarty33}. JPB and colleagues reached the conclusion that the GPT definitely supersedes the resonant model  \cite{Regnaut83} and demonstrated the importance of $s$-$d$ hybridization and overlap contributions to the effective interaction in noble metals \cite{Regnaut85} --- see also the discussion by Hausleitner et al.  \cite{Hausleitner1991}. This importance was also underlined in some studies based on the tight binding method \cite{TB1,TB2}.

\begin{figure}[!t]
\centering
\begin{minipage}{0.49\textwidth}
\begin{center}
\includegraphics[width=0.85\textwidth]{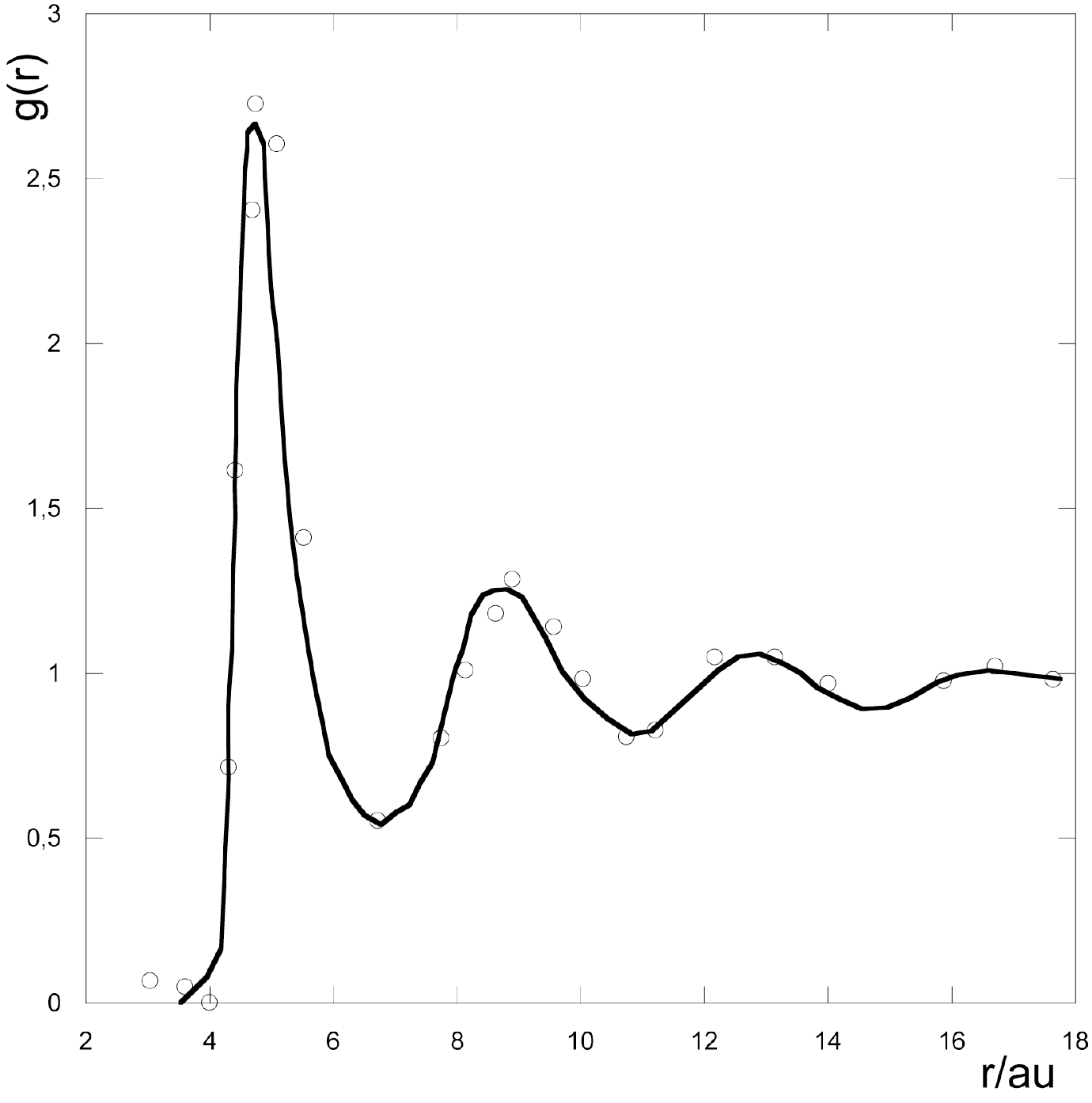}
\end{center}
\end{minipage}
\begin{minipage}{0.49\textwidth}
\begin{center}
\includegraphics[width=0.99\textwidth]{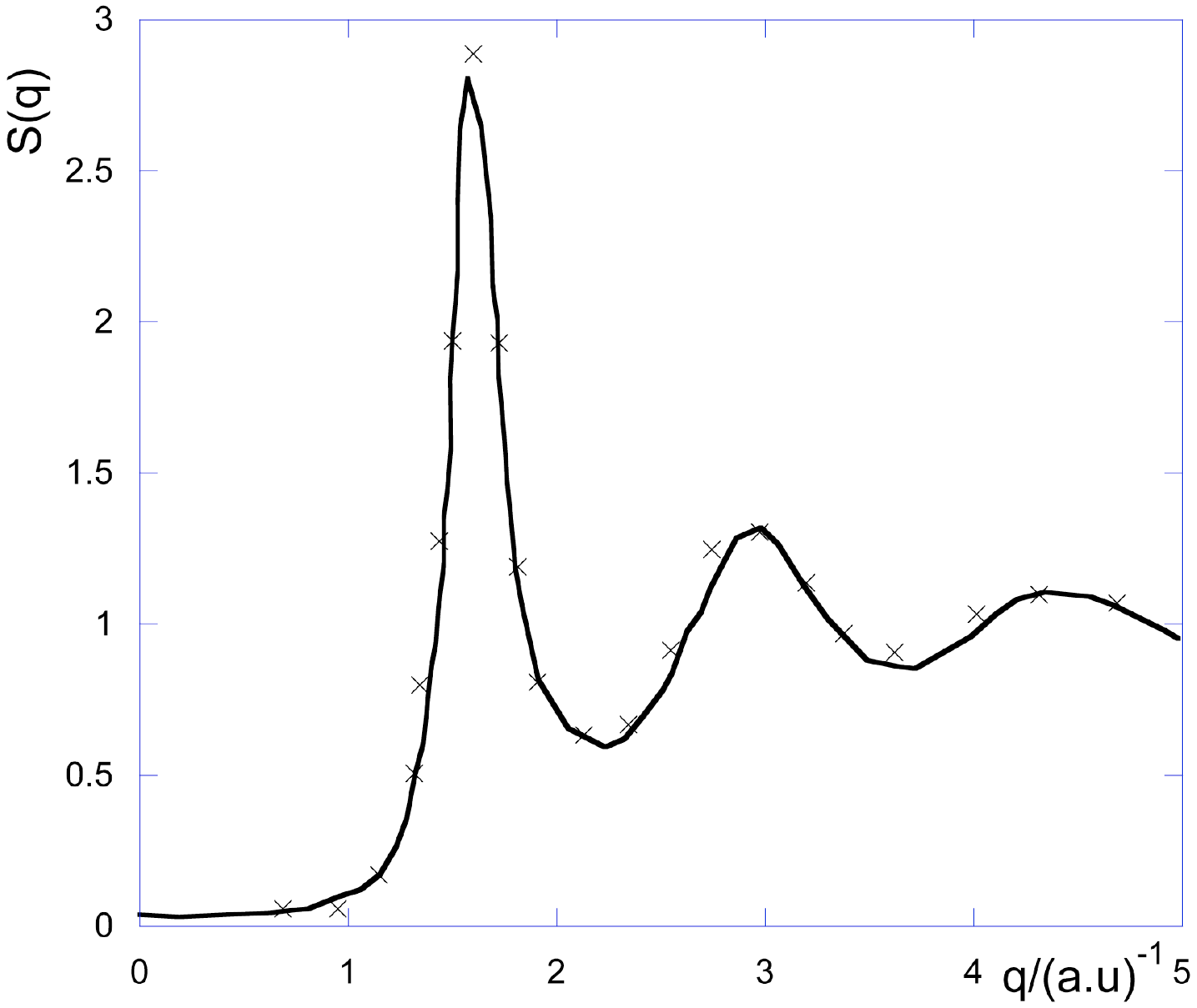}
\end{center}
\end{minipage}
\begin{minipage}{0.49\textwidth}
\begin{center}
\caption{Rdf of pure Cu near melting.
Symbols: xrd data of  \protect \cite{Waseda},  full curve: parametrized GPT \protect \cite{Regnaut83}. Distances in atomic units (0.52917~\AA.) }
 \label{grCu-Christ}
\end{center}
\end{minipage}
\begin{minipage}{0.49\textwidth}
\begin{center}
\caption{Structure factor of pure Cu at T $=1356$~K.
Symbols: xrd data of  \protect \cite{Waseda},  full curve: parametrized GPT \protect \cite{Regnaut83}. }
 \label{SqCu-Christ}
\end{center}
\end{minipage}
\end{figure}

\begin{figure}[!t]
\centering
\includegraphics[clip,width=0.5\textwidth]{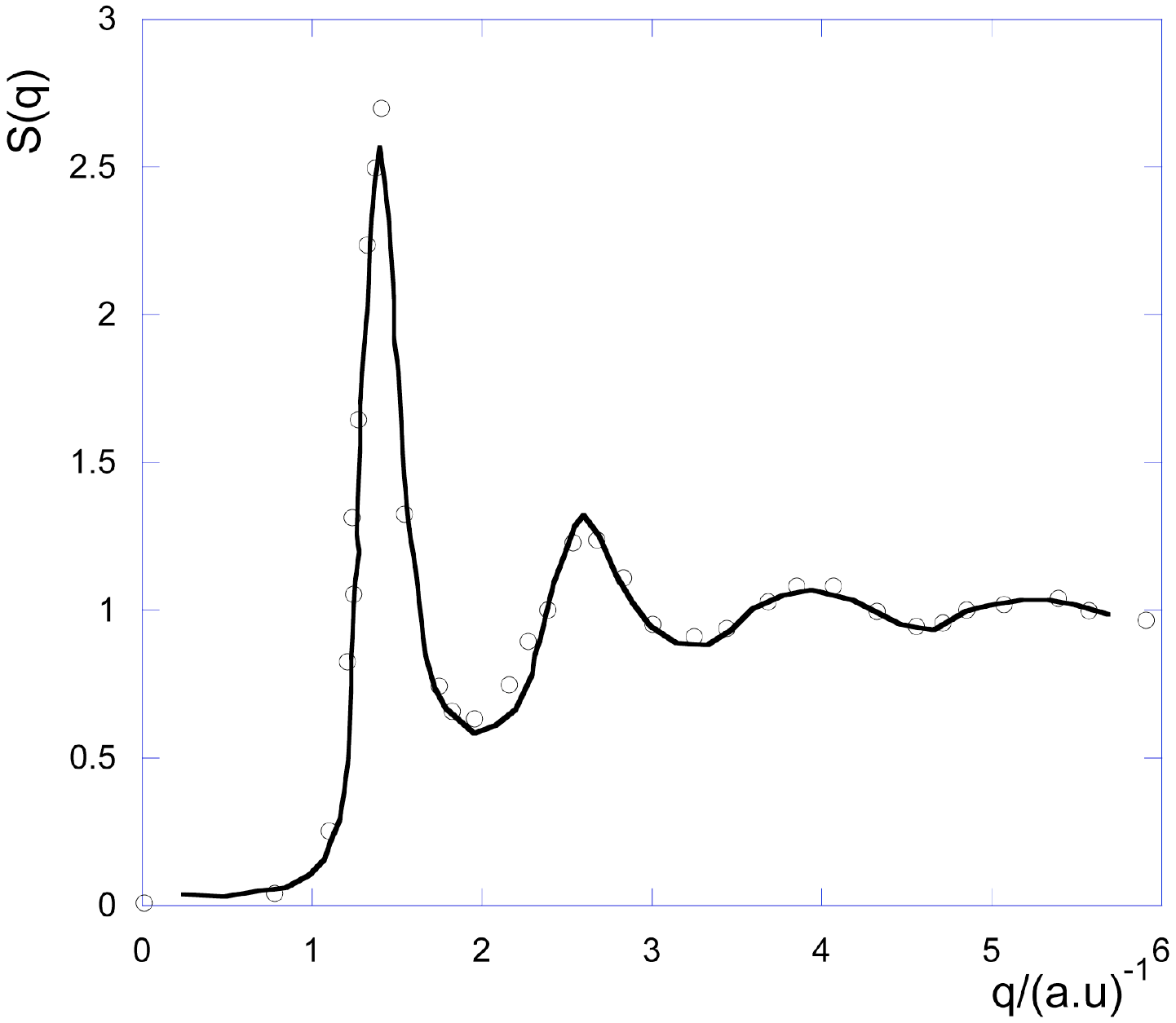}
\caption{Structure factor of pure Ag at T $=1234$~K.
Symbols: xrd data of  \protect \cite{Waseda},  full curve: parametrized GPT \protect \cite{Regnaut83}. }
\label{SqAg-Christ}
\end{figure}

From  a parametrized GPT theory, it was found  that  the liquid structure of copper is very well described as well as its resistivity  and thermodynamical properties.  This is illustrated in figures~\ref{grCu-Christ} (rdf)  and \ref{SqCu-Christ} (structure factor). The same approach fairly well describes silver as shown in figure~\ref{SqAg-Christ}, but it fails for Au \cite{Regnaut83,Fusco-these}.  The GPT theory was later  extended \cite{Moriarty1990} to  transition metals with partially filled $d$-bands. To our knowledge, this approach has not yet been used to study the liquid structure. By contrast, parametrized model potentials such as the Wills-Harisson one allow \cite{Wills}  a simple analysis of some trends in the liquid transition metals \cite{Regnaut89,Hafner-Hausleitner,Hafner-Hausleitner2}. Incidentally, using the effective number of $d$ electrons $z_d$ of the Wills-Harrisson model improves the calculation of  the work function of noble metals \cite{Russier88}.

In the last years, the GPT has been considerably improved, by including three  and four-body forces. A high degree of accuracy is thus achieved  for all transition metals \cite{Moriarty1990,Moriarty1997,Moriarty2006}. However, the same period witnessed the rise of  the AIMD approach, which, independently of the nature of the metal,  bypasses the separation of the total energy $E(\textbf{R}_1,\ldots,\textbf{R}_N)$ into a sum of $n$-body terms. This point is briefly discussed below (see also the discussion in figure~1 in \cite{Moriarty2006}).
\newpage

\subsection{Alloys} \label{sec5}
\subsubsection{Liquid alloys}

With mixtures of simple metals, some success has been achieved with interatomic potentials. Numerous studies have been performed  to correlate  the short and medium range structuring in the liquid  and the  phase diagram to the  effective pair potentials. Much has been done, first in the  simple case of the alkali metals (see \cite{Hafner87,Kahl_Hafner87,Kahl_1990,Hoshino,Canales,Wax}  and references therein) and  mixtures with polyvalent metals (see, for instance, \cite{Gonzales}). These studies either used first principles \PP s or a local  model to calculate  partial liquid structure factors and predict,  for instance, the homocoordination  preference and demixing \cite{Takhtoukh}  --- for an overview see \cite{Singh_report}. However, the fact that the agreement can be  better with simple local models  than with non-local norm-conserving  \PP\   for pure metals and  alloys \cite{Fiolhais}  shows that the situation is not completely clear and raises questions about the use  of the corresponding effective pair potentials in classical molecular dynamics \cite {Hellal}.
In the same vein, we mention  an attempt to go beyond mixtures of simple metals in \cite{ Khaleque}. On the other hand, the intricacy of the GPT theory for transition metals with partly filled $d$-band makes its implementation difficult, especially for alloys, though several studies have been done in this framework  \cite{Moriarty1990,Moriarty1997,Moriarty2006}.

In the late nineties, numerous AIMD studies of multicomponent alloys of simple metals, less simple  and transitions ones were conducted by different groups. A review of this work is beyond the scope of this contribution --- see for example \cite{Costa,Senda,Gonzales2,Jakse2008} and references therein. In the last section, we discuss some problems raised by  such simulations.

\subsubsection{Bulk metallic glasses}

In this last section, we illustrate, based on  our own work, one last issue in the study of liquid metals, namely the necessity for lengthy calculations to combine classical and \textit{ab initio} molecular dynamics.  The  specific \tern \ system we studied  indeed  combines several sources of difficulty: ternary alloy of transition metals, studied along a continuous path in the parameters space, with the underlying problem of the transferability of the semi-empirical pair potentials that are unavoidable, say, in a quench at a constant pressure.  This  system is a prototype of  \textit{bulk} glass forming alloys whose discovery in the late eighties \cite{Inoue1988} gave a new impetus to the study of amorphous metals.  They indeed contrast with the ordinary metallic glasses known from the early sixties \cite{Klement} that could be  elaborated only in the form of very thin ribbons and studied for decades now$^{1,2}$. Since then, such \textit{bulk} metallic glasses (BMG)  have been the subject of numerous studies, due to their unique physical, mechanical and corrosion properties \cite{Niinomi,Zhang_Plastic,Trexler,Gong}.

\begin{figure}[!t]
\centering
\includegraphics[clip,width=0.5\textwidth]{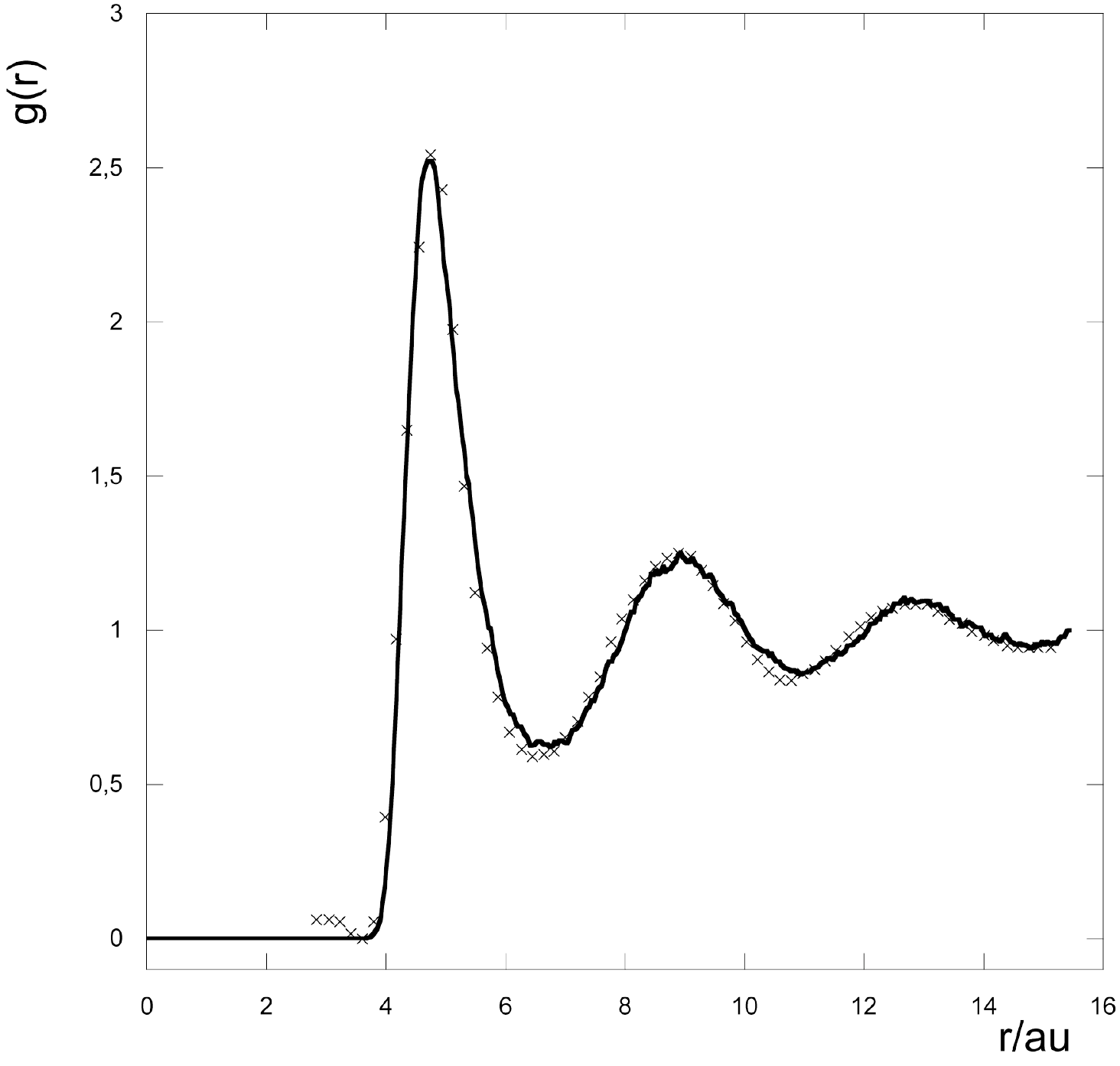}
\caption{Rdf of pure Cu at  $T=1635$~K.
Symbols: xrd data of  \protect \cite{Waseda} interpolated to 1635~K,  full curve: AIMD. }
  \label{gr_Cu-pur2}
\end{figure}

Actually,  a variety of combinations of AIMD and  CMD simulations have  been proposed \cite{FM1,Zope,Li,Tafipolsky,Jakse3}. Among these, the machine learning approach that uses neural networks \cite{Behler2007,Eshet,Behler,Artrith} to generate, given a training set of configurations, the weights for the  potential energy surface for  CMD simulations seems promising.   This method  may, however,  require  significant computer resources if the training is made on a large number of AIMD configurations.   We illustrate here  a particular combination of AIMD and  CMD  that makes possible  the (NPT) simulation of amorphization by melt quenching with moderate computer resources (24~cores in our recent work). As shown in \cite{SA2015} the basic idea is to use  AIMD  only at the initial and final points of the quench path. It proceeds as follows: (1) perform the \textit{ab initio} simulations to determine the rdfs $g^{\text{ab}}_{ij}(r)$ in the liquid (here $T=1600$~K); (2) determine the parameters denoted by  $\{a^l_i\}$  of the pair potential  by adjustment of the CMD $g_{ij}(r)$ on the  rdfs $g^{\text{ab}}_{ij}(r)$  at $T=1600$~K; (3) perform a quench down to $T=300$~K with the CMD simulation; (4) use  the last equilibrium configuration so obtained to run the \textit{ab initio} simulation at 300~K. The AIMD and CMD simulation were made with the Quantum Espresso (QE) \cite{QE} and LAMMPS \cite{LAMMPS} packages, respectively.  The   CMD simulations were  made  using the simplified (i.e., without 3-body terms) Stillinger-Weber \cite{SW} (SW) potentials used in \cite{Teichler1}.  The rdfs  of \text{\tern} were determined from  \textit{ab initio} simulation with ultra-soft pseudopotentials from the QE library.
$N=260$ particles ($156+52+52$ atoms for Cu, Zr, and Ti, respectively) were used, to keep the CPU time in a tolerable range on the  24-core computer used. In these conditions, 1500~steps of accumulation generate, in about 45~days of simulation, reasonably smooth curves for $g_\text{Cu-Cu}$, $g_\text{Cu-Ti}$, and $g_\text{Cu-Zr}$. With the good scalability with the number of processors of the PWscf code, this restitution time can be reduced to about one week, still with affordable computer resources. Due to  the low statistics for the minority species,   $g_\text{Ti-Ti}$, $g_\text{Ti-Zr}$,  and $g_\text{Zr-Zr}$ in \cite{SA2015} were quite noisy but no significant change was found after  additional accumulation --- up to 3500~steps --- made recently.
To  illustrate  the method for the pure liquids, the rdf of pure liquid copper is shown in figure~\ref{gr_Cu-pur2}. As already noticed in \cite{SA2015}, comparison with figure~\ref{grCu-Christ} shows that  the GPT theory does almost as well as the AIMD simulation with PAW \PP\ for pure~Cu.

\begin{figure}[!b]
\centering
\begin{minipage}{0.49\textwidth}
\begin{center}
\includegraphics[width=0.9\textwidth]{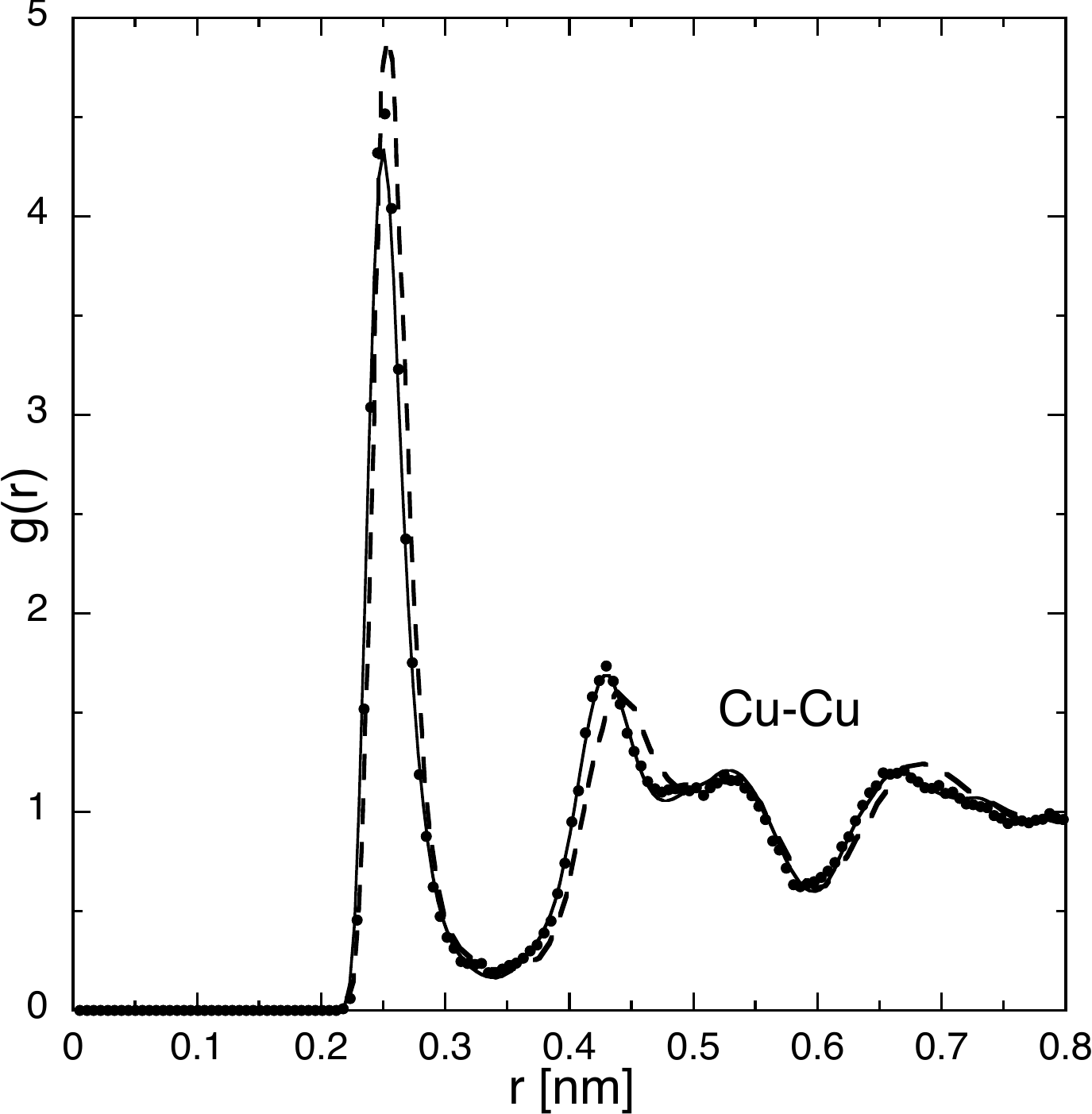}
\end{center}
\end{minipage}
\begin{minipage}{0.49\textwidth}
\begin{center}
\includegraphics[width=0.99\textwidth]{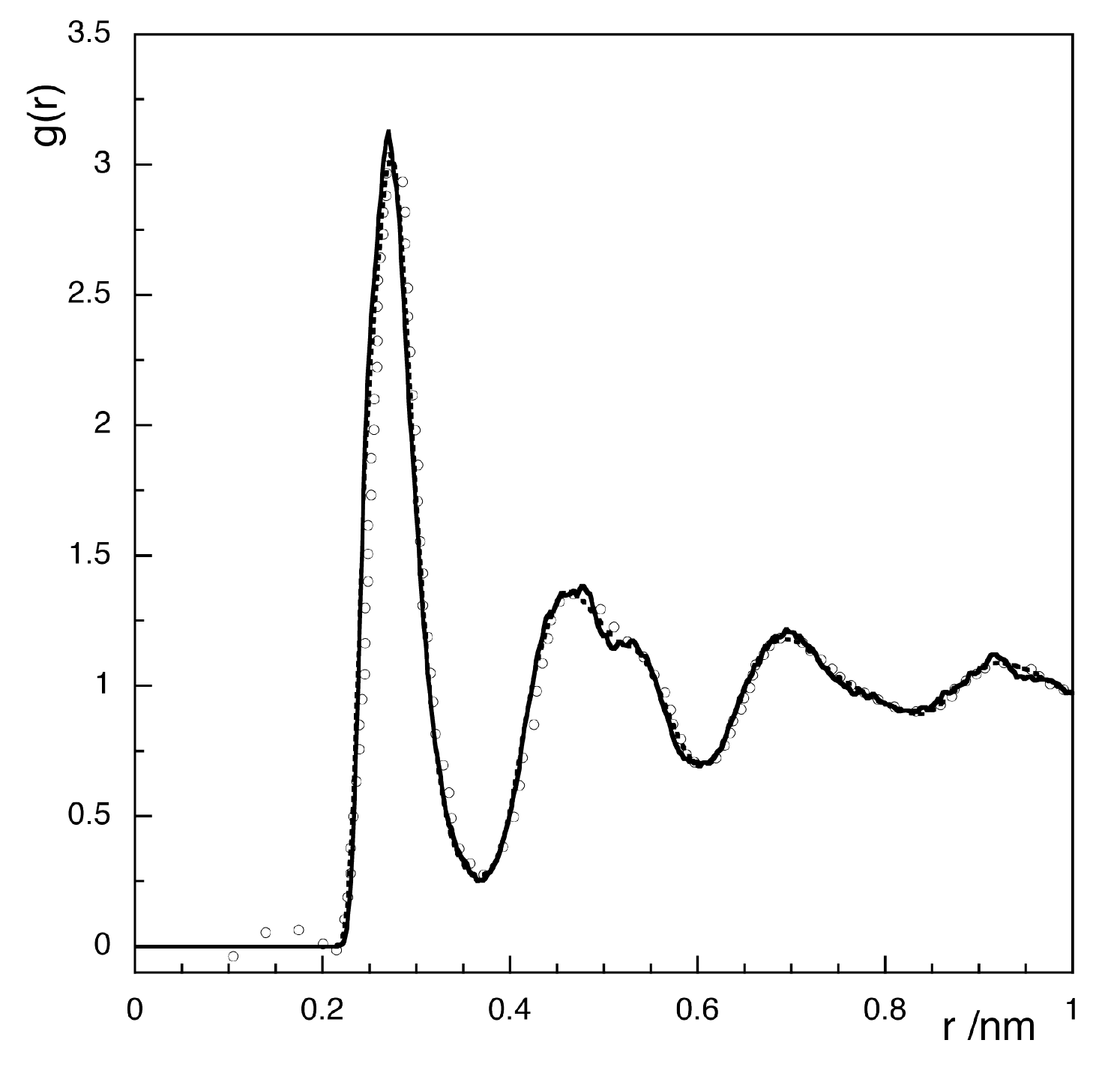}
\end{center}
\end{minipage}
\begin{minipage}{0.49\textwidth}
\begin{center}
\caption{Partial rdf of copper in  amorphous \tern  \ at $T=300$~K.
 Full curve: AIMD, dots: CMD with the parameters adjusted at 300~K, dashes: MD simulation \protect \cite{Dalgic} with tight-binding potentials for Cu$_{50}$Ti$_{25}$Zr$_{25}$. }
\label{grCu-Cu_amorph}
\end{center}
\end{minipage}
\begin{minipage}{0.49\textwidth}
\begin{center}
\caption{Experimental and simulated total rdf of amorphous \tern  \ at  $T=300$~K.
Symbols: xrd data of Durisin et al. \protect\cite{Durisin},  full curve: AIMD, dashes:  CMD with the parameters adjusted at 1600~K.
}
\label{grCu-Ti-Zr_amorph}
\end{center}
\end{minipage}
\end{figure}

In figure~\ref{grCu-Cu_amorph}, we show the partial  rdf for Cu in the ternary alloy after the quench down to $300$~K.  $g_{\text{Cu-Cu}}$ obtained with the SW potentials with the  parameters $\{a^l_i\}$ (those  adjusted at $1600$~K) are compared to the AIMD ones. The conclusion is that the parameters are here  transferable from the liquid to the amorphous alloy, which is an essential feature of the CMD approach with parametrized force fields. This is a considerable saving of computer time indeed. The direct comparison with experiment  made in figure~\ref{grCu-Ti-Zr_amorph} illustrates the efficiency of the AIMD which justifies the use of the designation of  in-silico experiments.

To summarize, the  results presented in this section illustrate the possibility  of a quantitative determination of the pair structure during the quench by  an appropriate combination of \textit{ab initio} and classical molecular dynamics simulations.They again underline the importance of developing thoroughly tested parametrized
force fields, in order to keep simulations convenient enough for a quantitative study of the behavior of complex materials.

\section{Conclusion}

In conclusion,  this overview of the work done during four decades, especially the studies that followed from  the initial impulse given by JPB, shows a considerable progress made in the search of quantitative methods for studying  metallic elements. A gradual switch from approximate semi-analytic methods to the accurate numerical treatment of the  quantum many-body electron-ion problem has been emphasized. Nowadays, the most efficient simulation methods seem to be the ones that use classical MD simulations whose  force fields  are parametrized using the data obtained by AIMD simulations. The resulting approach is flexible and accurate enough to enable a reliable  exploration of the parameters space, as required, for example, in the construction of phase diagrams. Our recent work in this domain pursued the hope of Jean-Pierre Badiali to describe complex and interfacial  systems starting from a true microscopic approach, with well controlled theoretical methods.

\newpage
\ukrainianpart

\title{Рідкі метали: ранні роботи та деякі недавні дослідження}
\author{К. Реньо, С. Амокран}
\address{Група ``Фізика рідин і складних середовищ'',  факультет природничих і технічних наук, Університет Парі-Ест, Франція}

\makeukrtitle

\begin{abstract}
В цій статті ми обговорюємо розвиток теоретичних досліджень рідких металів, виконаних за останні десятиліття, починаючи з прикладу рідкого галію і ранніх робіт в групі Жана-П'єра Бадіалі. Вони грунтуються на поєднанні теорії збурень з псевдопотенціалами для електронів і теорії рідкого стану для іонів. Наприкінці, на прикладі склоутворюючих розплавів, проілюстровано пізніші дослідження, які поєднують метод \textit{ab initio} та класичний метод молекулярної динаміки.

\keywords рідкі метали, структура, теорія збурень, моделювання
\end{abstract}
\end{document}